\documentclass{PoS}

\def\gappeq{\mathrel{\rlap {\raise.5ex\hbox{$>$}}
{\lower.5ex\hbox{$\sim$}}}}
\def\lappeq{\mathrel{\rlap{\raise.5ex\hbox{$<$}}
{\lower.5ex\hbox{$\sim$}}}}

\title{Gravitino Dark Matter with Sneutrino NLSP in NUHM (15'+5')}

\ShortTitle{Gravitino Dark Matter with Sneutrino NLSP in NUHM (15'+5')}

\author{\speaker{Yudi SANTOSO} %
         \thanks{IPPP/08/79, DCPT/08/158}\\
        IPPP, University of Durham, UK\\
        E-mail: \email{yudi.santoso@durham.ac.uk}}

\abstract{The identity of dark matter has not been solved up to this date, 
          a problem that became the main topic of this conference. 
          There are many theoretical candidates for dark matter particle, including gravitino from supergravity models. For gravitino dark matter scenario, the phenomenology depends much on what the next lightest supersymmetric particle (NLSP) is. We show here that sneutrino can naturally be the NLSP in the Non-Universal Higgs Masses (NUHM) model, 
          and that this scenario is still phenomenologically viable.}

\FullConference{Identification of dark matter 2008\\
                 August 18-22, 2008\\
                 Stockholm, Sweden}

\begin{document}

\section{Introduction}

In supergravity models we require a spin 3/2 particle called gravitino. Gravitino as a candidate for dark matter is an interesting possibility that has regained more attention in the recent years. To be the dark matter  gravitino needs to be stable, and this can be realized if gravitino is the lightest supersymmetric particle (LSP) and the R-parity which prevent the LSP from decaying to standard model particles is conserved. Light gravitino is natural in gauge mediated supersymmetry breaking (GMSB) models~\cite{Giudice:1998bp}. In gravity mediated supersymmetry breaking models (SUGRA), the gravitino mass is naturally about the same order as the other soft masses and therefore could also be smaller (see e.g.~\cite{Nilles:1983ge}). We will assume gravity mediated models with weak scale SUSY masses and O(1-100~GeV) gravitino mass, treated as a free parameter, for the rest of this paper.  

Gravitino is a very weakly interacting particle. Its coupling to fermion fields is suppressed by the Planck mass, i.e. $\sim 1/M_{\rm Pl}$~\footnote{This is for gravitational interaction in SUGRA models. In GMSB the coupling is $k/F$ where $k < 1$ and $F$ is the mass splitting parameter.}. As a consequence, the NLSP is typically long lived, and this is particularly true for models with conserved R-parity in which the NLSP can decay only to the gravitino LSP. Another consequence of this very weak interaction is that aside from its gravitational effect, gravitino is practically undetectable. Therefore the discovery in this scenario relies on the search of the NLSP. At colliders, any supersymmetric particles produced would then quickly decay to the NLSP which would appear as a massive stable particle due to its long lifetime.

There are many possibilities for the NLSP, and each has its own distinct phenomenology. Here we focus on just one case, i.e. the `left-handed' sneutrino NLSP~\cite{Ellis:2008as}. The outline is as follows: in section~2 we discuss the phenomenological constraints of our scenario, in section~3 we discuss the NUHM realization, and then we summarize in section~4.

\section{Phenomenological Constraints}

\begin{figure}
\begin{center}
\includegraphics[width=0.48\linewidth]{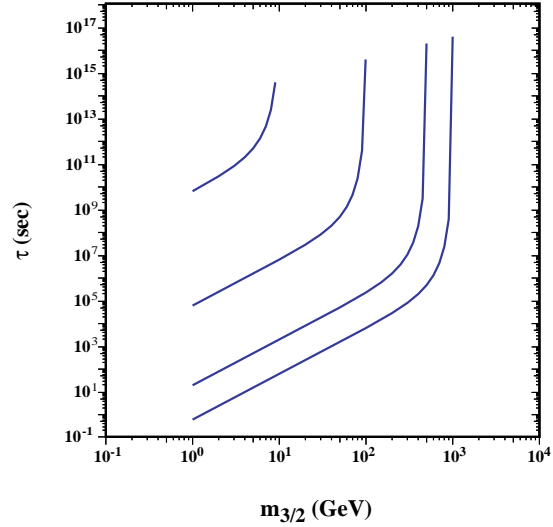}
\end{center}
\vskip -0.2in
\caption{Sneutrino lifetime as a function of gravitino mass $m_{3/2} \equiv m_{\tilde{G}}$, for various values of sneutrino mass, $m_{\tilde{\nu}} = 10, 100, 500, 1000$~GeV, respectively from left to right. \label{fig:lifetime}}
\end{figure}

It has been understood that if there is a metastable particle that decays during or after the Big Bang Nucleosynthesis (BBN) era, then the primordial light element abundances could be different from the standard BBN (SBBN) theory prediction. The fact that SBBN is a highly succesful theory put a strong constraint on any theory with metastable particle including the gravitino dark matter scenario that we are considering here. So first we need to calculate the sneutrino lifetime. It is clear that the dominant decay channel of sneutrino is the 2-body decay $\tilde{\nu} \to \tilde{G} + \nu$, with decay rate
\begin{equation}
\Gamma_{\rm 2b} = \frac{1}{48 \pi} 
\frac{m_{\tilde{\nu}}^5}{M_{\rm Pl}^2 m_{\tilde{G}}^2} 
\left( 1 - \frac{m_{\tilde{G}}^2}{m_{\tilde{\nu}}^2} \right)^4 
\end{equation}
where $m_{\tilde{G}}$ is the gravitino mass, $m_{\tilde{\nu}}$ is the sneutrino mass, and $M_{\rm Pl}$ is the reduced Planck mass. Therefore we can determine sneutrino lifetime as $\tau \simeq 1/\Gamma_{\rm 2b}$. The numerical result is presented in Fig.~\ref{fig:lifetime}. We see that the lifetime has a very large range depending on the relative values of $m_{\tilde{G}}$ and $m_{\tilde{\nu}}$. The sneutrino can be very stable if $(m_{\tilde{\nu}} - m_{\tilde{G}})/m_{\tilde{G}} \ll 1$. We focus our attention to the case with $\tau \sim (1 - 10^6)$~s, in which the strongest constraint comes from BBN. 

The effects of metastable particle on BBN come in two categories: electromagnetic and hadronic. Since our NLSP, the sneutrino, is neutral we might think that its BBN effect should be negligible. However, it was pointed out in~\cite{Kanzaki:2007pd} that there could still be significant effect: from energy transfer from the energetic neutrinos produced by the decay to the ambient charged particles; and also from the $3$- and $4$-body decays (i.e. $\widetilde{\nu} \to \widetilde{G} + \nu + (\gamma, Z)$,  $\widetilde{\nu} \to \widetilde{G} + \ell + W$ and $\widetilde{\nu} \to \widetilde{G} + \nu + f + \bar{f}$,
$\widetilde{\nu} \to \widetilde{G} + \ell + f + \bar{f}^\prime$, respectively). Eventhough the branching ratios for these decays are small, they could produce some energetic quarks which lead to hadronic showers directly.

Aside from the 3- and 4-body branching ratios, an important factor that determine the extent of the effects is of course the amount of the decaying sneutrino. The sneutrino decoupled from the thermal plasma before it decayed, therefore the freeze out density can be calculated by the
usual method of solving the Boltzmann equation in the expanding Universe.  
The yield, $Y_{\tilde{\nu}} = n_{\tilde{\nu}}/s$, is then determined by 
\begin{equation}
Y_{\tilde{\nu}} m_{\tilde{\nu}} = \Omega_{\tilde{\nu}} h^2 \times (3.65 \times 10^{-9} \; {\rm GeV}) 
\end{equation}

The sneutrino mass is constrained from below by the $Z$ boson decay width measurement by LEP to be $m_{\tilde{\nu}} \gappeq 43$~GeV~\cite{Ellis:1996xu}. At colliders, all other sparticles produced would quickly decay to sneutrino which would then escape the detector as missing energy. 
A more stringent limit on the sneutrino mass would need detail analysis using models with stable sneutrino, which might yield different result from that with neutralino LSP assumption as quoted in PDG~\cite{Amsler:2008zz}. 


\section{Sneutrino NLSP in NUHM}

\begin{figure}
\begin{center}
\includegraphics[width=0.50\linewidth]{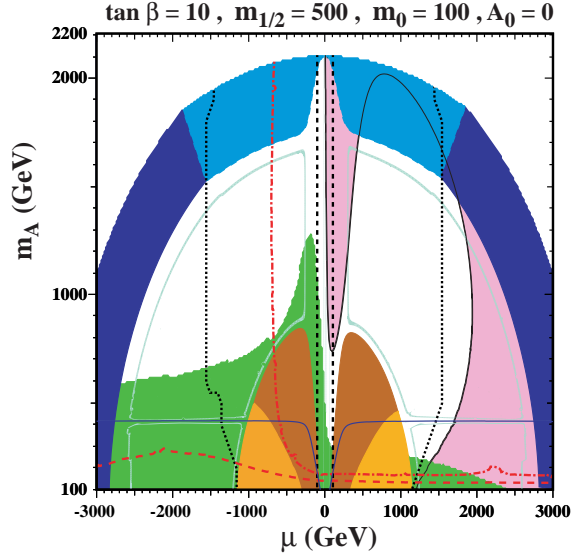}
\end{center}
\vskip -0.2in
\caption{The $\mu - m_A$ plane in NUHM for $\tan \beta = 10$, $m_{1/2} = 500$~GeV, $m_0 = 100$~GeV, and $A_0 = 0$. \label{fig:mumA}}
\end{figure}

The Non-Universal Higgs Masses (NUHM) model is a supersymmetric model with six SUSY parameters: the universal scalar soft mass $m_0$, universal gaugino mass $m_{1/2}$, and universal trilinear coupling $A_0$, all defined at the GUT scale; the ratio of the two Higgs vev $\tan \beta$, the Higgs mixing parameter $\mu$ and the CP odd Higgs mass $m_A$. In this model the soft masses for the Higgs doublets are, in general, not equal to $m_0$ at the GUT scale. 

This non-universality of the Higgs masses has non-vanishing effect on the RGE evolution of the sfermion masses. The crucial observation is that the non-universal terms has opposite signs between the left and right `handed' sleptons RGE. If the non-universality is large we could have `left handed' slepton lighter than the `right handed' one, in contrast to the universal case where the right handed sleptons are always lighter because their masses do not get contribution from the $SU(2)$ interaction. Furthermore, there are also $D$-term contributions to the slepton masses, which split the charged-slepton from the sneutrino in the left doublet. For relatively large $\tan \beta$, sneutrino is the lighter. By choosing large $\mu$ and/or large $m_A$, we found that sneutrino could be the lightest among the MSSM sparticles. This is shown in Fig.~\ref{fig:mumA}, the blue-shaded region is the sneutrino region. The non-universality also goes into the Yukawa terms in the RGE, and because of this, we can have either the third generation sneutrino or the first two generations be the lighter. The dark blue in Fig.~\ref{fig:mumA} represents $\tilde{\nu}_{e,\mu}$ regions, while the light blue represents  $\tilde{\nu}_\tau$ regions. 
   
\begin{figure}
\begin{center}
\vskip -0.1in
\includegraphics[width=0.40\linewidth]{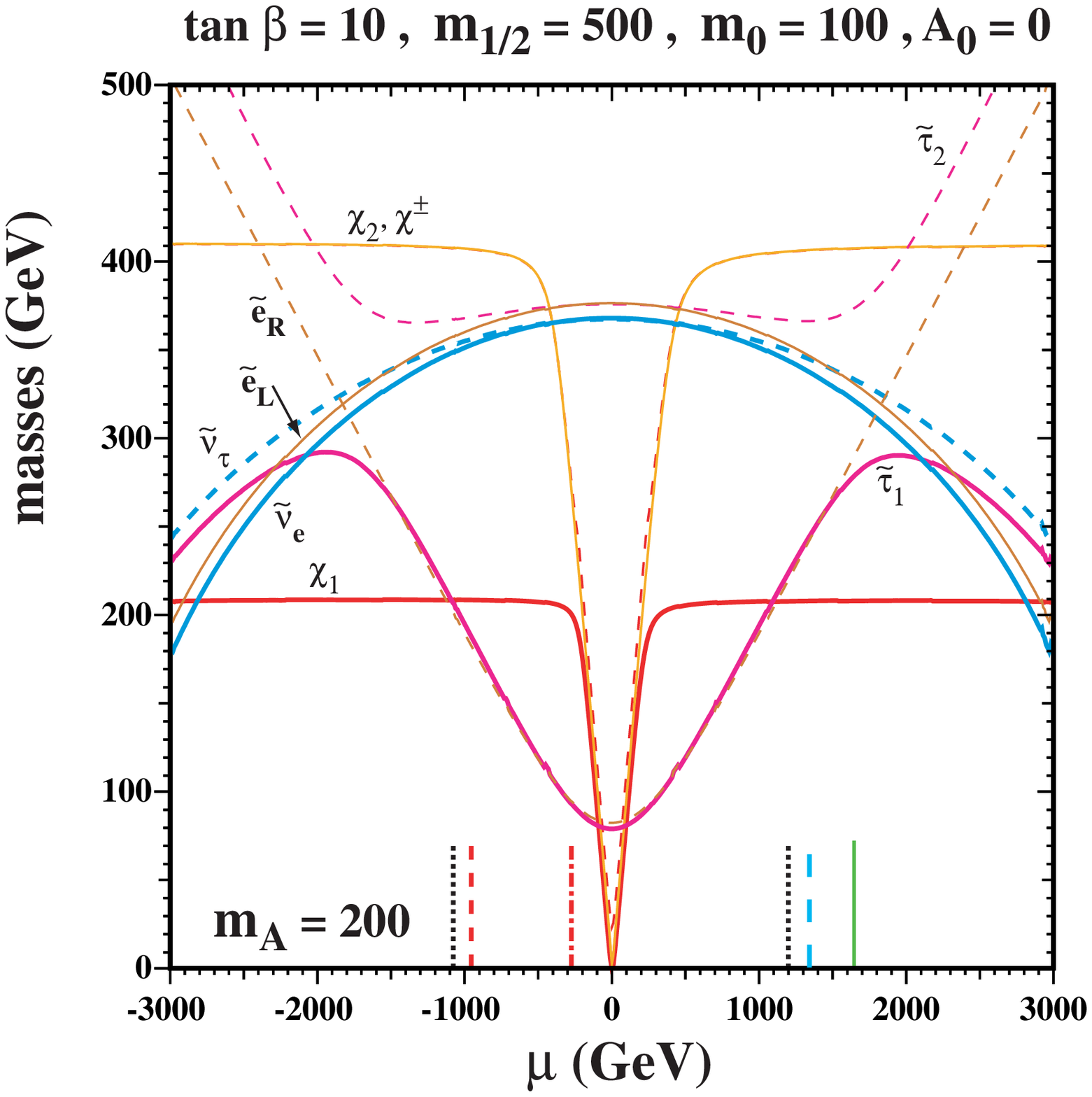}
\hspace{0.5cm}
\includegraphics[width=0.40\linewidth]{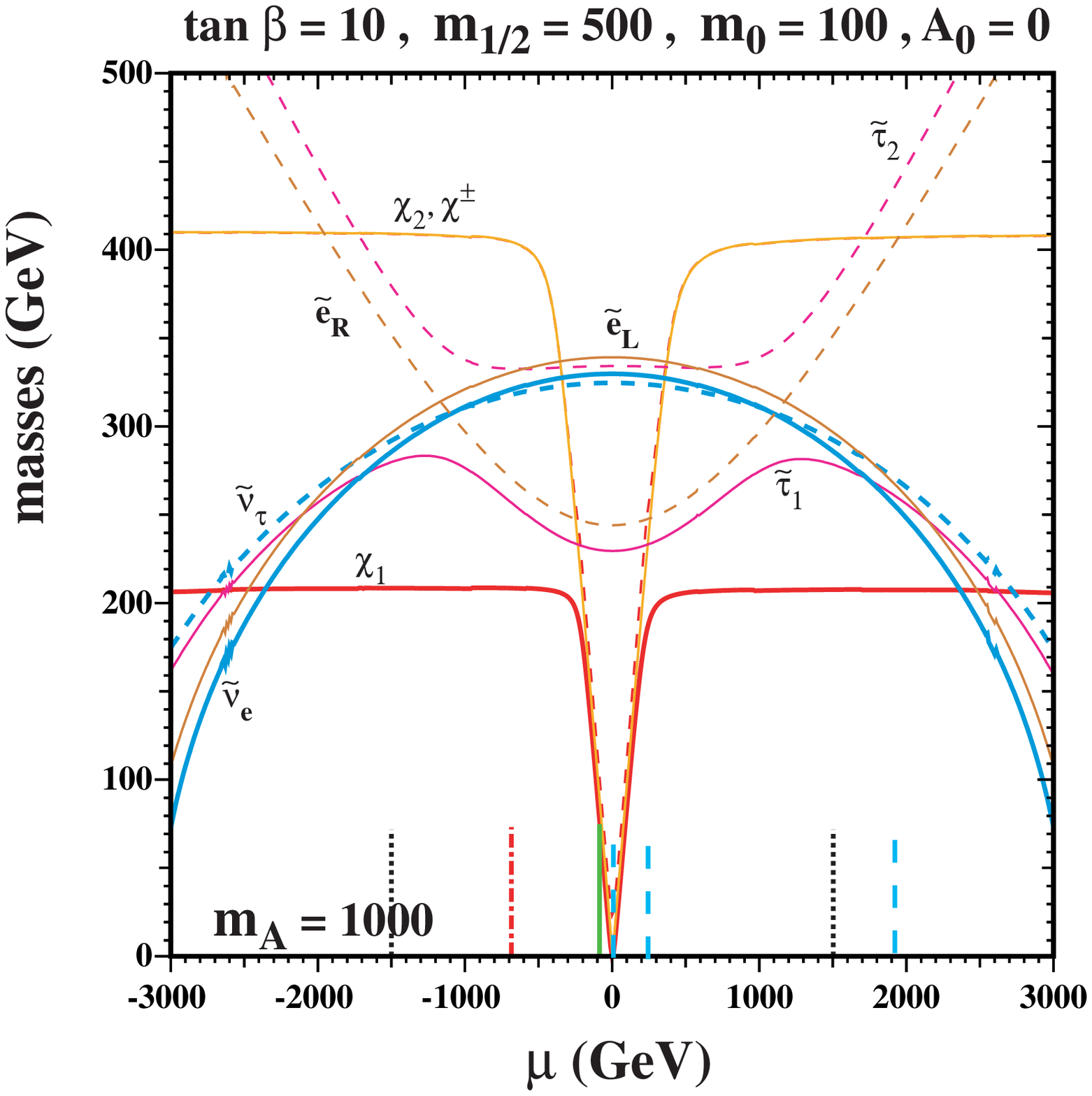}
\vskip 0.02in
\includegraphics[width=0.40\linewidth]{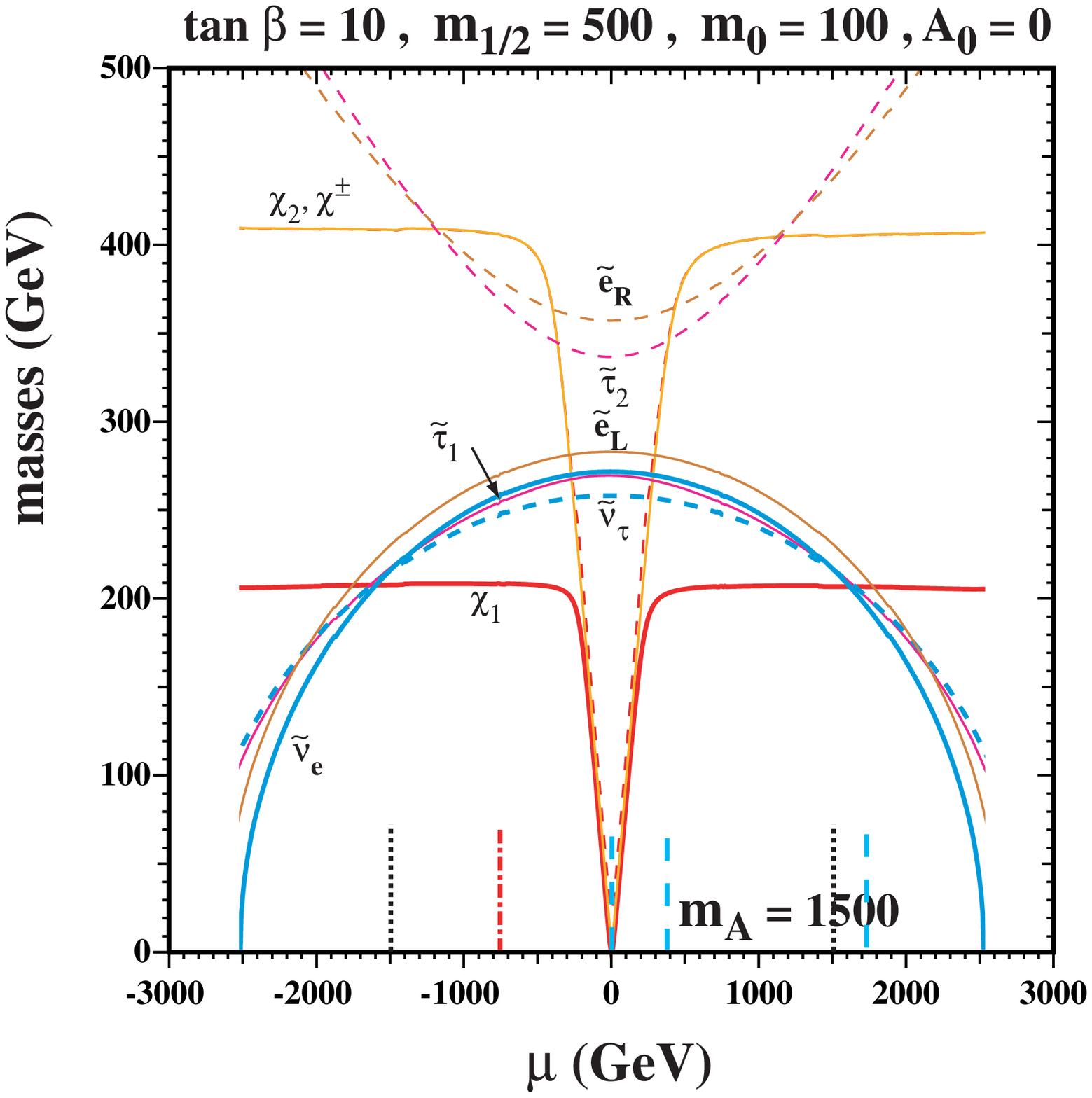}
\hspace{0.5cm}
\includegraphics[width=0.40\linewidth]{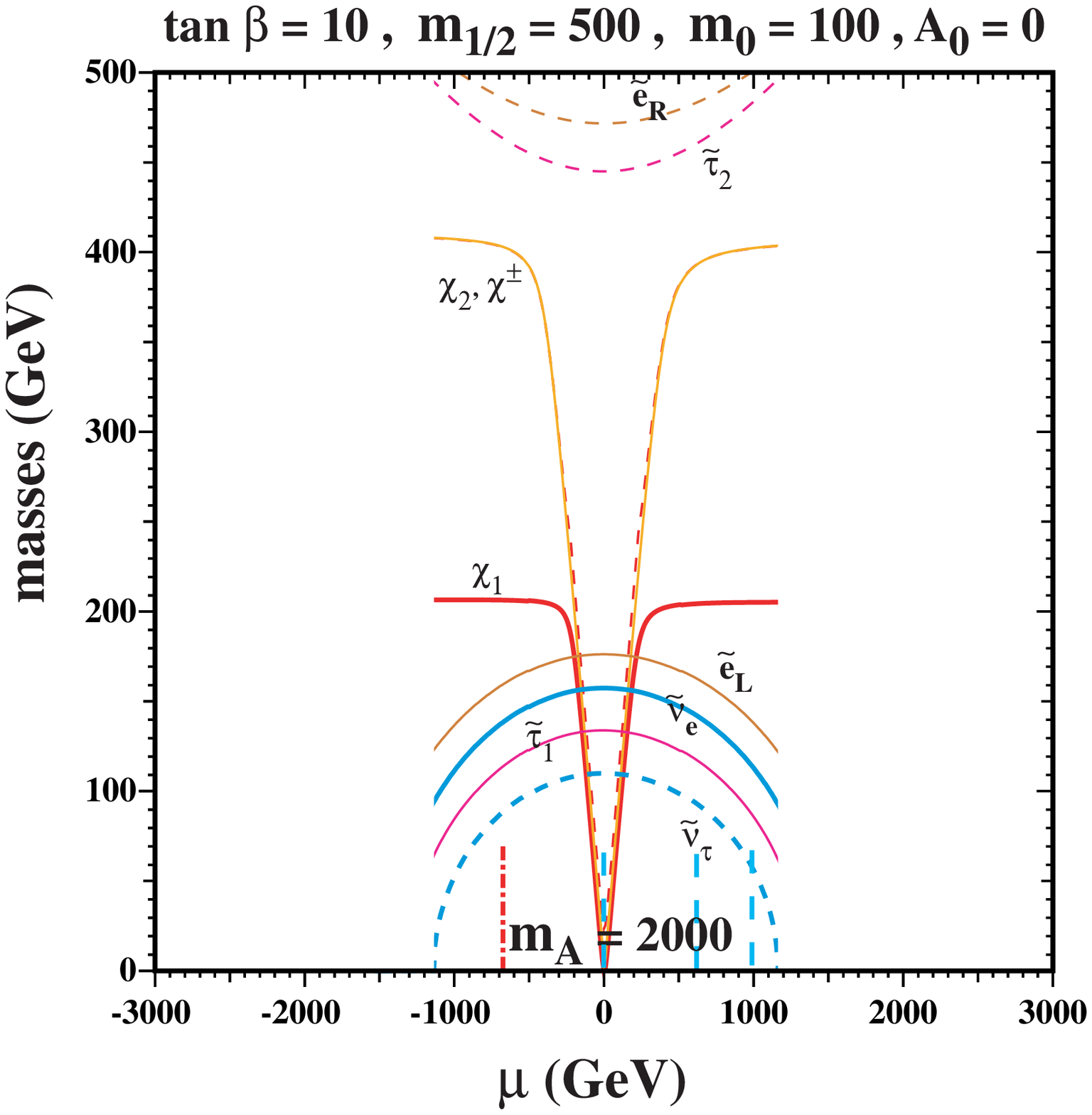}
\end{center}
\vskip -0.25in
\caption{Sparticle masses as functions of $\mu$, for $\tan \beta = 10$, $m_{1/2} = 500$~GeV, $m_0 = 100$~GeV, $A_0 = 0$ and $m_A = 200, 1000, 1500$ and 2000~GeV, respectively. \label{fig:masses}}
\end{figure}

We show in Fig.~\ref{fig:masses} the sparticle masses as functions of $\mu$ for  certain values of $m_A$,  corresponding to horizontal slices of Fig.~\ref{fig:mumA}. Notice that for regions with sneutrino as the lightest, there are some other sparticles (i.e. sleptons) which are not much heavier than the sneutrino. This structure originates from the universality assumption of the scalar masses at the GUT scale.

We then calculate the yield and the branching ratios for the sneutrino NLSP in the allowed regions. For the parameter space that we considered, we found that the 3-b branching ratio is less than $\sim 10^{-6}$ and the yield $Y_{\tilde{\nu}} m_{\tilde{\nu}} \lappeq 10^{-11}$~GeV. From comparing with the results in~\cite{Kanzaki:2007pd}, we conclude that our models are not excluded by the BBN constraint.

\section{Summary}

We have discussed the sneutrino NLSP scenario with gravitino dark matter assumption within the NUHM framework. We found that this scenario is still allowed by all the known constraints. The upcoming LHC experiment at CERN would be able to tell us more about supersymmetry. Given that neutralino LSP is not the only possibility, a detail study of the collider signatures for this scenario, following the preliminary study of~\cite{Covi:2007xj}, would be justifiable.

\section*{Acknowledgments}

I would like to thank John Ellis and Keith Olive for collaboration on the research presented in this talk; and the conference organizer for such a wonderful and fruitful conference. I am pleased to acknowledge useful conversations with Bhaskar Dutta, Terrance Figgy and Jihn~E.~Kim.

\end{document}